\documentstyle[12pt,epsf]{article}
\begin{document}

\title{Non-Abelian Stokes Theorem and Computation of Wilson Loop}
\author{Ying Chen${}^{2}$\thanks{Email address: cheny@hptc5.ihep.ac.cn},
         Bing He${}^{2}$\thanks{Email address: heb@alpha02.ihep.ac.cn},
Ji-min
Wu${}^{1,2}$\thanks{Email address: wujm@alpha02.ihep.ac.cn}\\
{\small ${}^{1}$ CCAST (World Laboratory), P.O.Box 8730, Beijing
100080, P.R.China}\\
   {\small ${}^{2}$ Institute of High Energy Physics, Academia Sinica,
Beijing
100039, P.R.China}}

\maketitle

\vspace{4mm}
\begin{abstract}
It is shown that the application of the non-Abelian Stokes theorem to the
computation
of the operators constructed with Wilson loop will lead to ambiguity, 
if the gauge field under consideration is a non-trivial one. 
This point is illustrated by the specific
examples of the computation of a non-local operator.
\noindent
\end{abstract}
\baselineskip 0.3in
\vskip 0.5cm
The non-Abelian Stokes theorem$^{1-5}$ is widely applied to compute
Wilson loop (closed non-Abelian phase factor), $\Psi(C)=Pexp\left(ig\oint_C 
dz^{\mu}A_{\mu}(z)\right)$ 
($A_{\mu}$ is the simplified notation
for $\sum\limits^{dimG}_{a=1}A_{\mu}^aT^a$), 
which is important for the construction of 
gauge invariant operators in the non-perturbative approaches to QCD.
The power of the theorem lies in transforming the line integral in
a Wilson loop to a more tractable surface integral over the surface $S$
enclosed by contour
$C$: 
\begin {eqnarray}
Pexp\left(ig\oint_Cdz^{\mu}A_{\mu}(z)\right)=\Psi(x,y;\bar C_x)
Pexp\left(ig/2\int_S 
d\sigma^{\mu\nu}\hat{F}_{\mu\nu}(y,z;C_z)\right)\Psi(y,x;C_x),
\end{eqnarray}
where $y$ is an arbitrary reference point on $S$, $\Psi(y,x;C_x)$ the
phase
factor connecting the initial and final point $x$ of $C$ to $y$
through a path $C_x$ and its inverse path $\bar C_x$. 
The shifted field strength $\hat{F}_{\mu\nu}(y,z;C_z)$ here is defined
as $\hat{F}_{\mu\nu}(y,z;C_z)=\Psi(y,z;C_z)F_{\mu\nu}(z)\Psi^{\dag}(y,z;C_z)$,
and its ordered integral over $S$ is given as follows:
\begin{eqnarray}
Pexp\left(ig/2\int_Sd\sigma^{\mu\nu}\hat{F}_{\mu\nu}(y,z;C_z)\right)
~=\lim_{n\rightarrow\infty}\prod\limits_{i=n}^{1}(I+
ig/2\hat{F}_{\mu\nu}
(y,z_i;C_{z_i})\Delta\sigma^{\mu\nu})
\end{eqnarray}
$$=I+ig/2\int^1_0ds\int^1_0dt\frac{\partial(z^{\mu},z^{\nu})}
{\partial(s,t)}\hat{F}_{\mu\nu}(y,z(s,t))+
(ig/2)^2\int^1_0ds\int^1_0dt
\frac{\partial(z^{\mu},z^{\nu})}{\partial(s,t)}
\hat{F}_{\mu\nu}(y, z(s,t))$$
\begin{eqnarray}
\times\int^1_0ds_1\int^t_0dt_1\frac{\partial(z_1^{\mu},
z_1^{\nu})}{\partial(s_1,t_1)}\hat{F}_{\mu\nu}(y,z_1(s_1,t_1))+\cdots.
\end{eqnarray}
Eq. (2) means that the ordered surface integral is equivalent to the
infinite product
of the phase factor on the net of the handled
small plaquettes around each $z_i$ in $S$. Obviously the `handle'
is referred to as the phase factor connecting $z_i$ and $y$: 
$$(I+ig/2\hat{F}_{\mu\nu}(y,z_i;C_{z_i})\Delta\sigma^{\mu\nu})
=\Psi(y,z_i;C_{z_i})(I+ig/2F_{\mu\nu}(z_i)\Delta\sigma^{\mu\nu})
\Psi^{\dag}(y,z_i;C_{z_i}).$$
If the shifted field strength $\hat{F}_{\mu\nu}(y,z; C_z)$ is well defined
as the function of $\{s,t\}$ in the
equations, the ordered surface 
integral in Eq. (2) can be performed definitely, and it makes the 
calculation of the
expectation value $<Tr\Psi(C)>$ convenient by means of the cumulant 
expansion 
technique$^6$. For an overview see Ref. [7].  
However, we find that the transformation from the line integral to
the surface integral in Eq. (1) might also give rise to an ambiguity
in the computation of Wilson loop $\Psi(C)$, 
if it is applied in the situation of a non-trivial gauge field. In this letter 
we will present the specific     
examples to illustrate the point. For simplicity we specify that 
the gauge field under consideration is two-dimensional and the gauge group 
is SU(2).   

Before our discussion we review some properties of the non-Abelian
phase factor $\Psi(x,x_0;C_x)$ that is defined as the solution
of the following differential equation:
\begin{eqnarray}
\frac{d\Psi}{dt}(t)=A(t)\Psi(t),
\end{eqnarray}
with the boundary condition
$$\Psi(t_0)=I,$$
where $A(t)=igA_{\mu}(x(t))dx^{\mu}(t)/dt$.
The path $C_x$ is parametrized by $t$, with $x(t)=x$ and $x(t_0)=x_0$.
The solution of Eq. (4) is 
$$\Psi(t)=
Pexp\left(\int^{t}_{t_0}dsA(s)\right)~=I+\int^{t}_{t_0}dsA(s)$$
$$~+\int^t_{t_0}dsA(s)\int^{s}_{t_0}ds_1A(s_1)+\cdots+
\int^t_{t_0}dsA(s)\cdots\int^{s_{n-1}}_{t_0}ds_{n}A(s_{n})+\cdots$$
\begin{eqnarray}
=\lim_{\Delta t \rightarrow 0}(I+A(t_{n-1})\Delta t)\cdots 
(I+A(t_{0})\Delta t). 
\end {eqnarray}
From it we immediately obtain the following two properties of 
$\Psi(x,x_0;C)$:
\begin{eqnarray}
\Psi(x_3,x_1; C_{x_2}\circ C_{x_1})=\Psi(x_3,x_2;C_{x_2})
\Psi(x_2,x_1;C_{x_1}); 
\end{eqnarray}
\begin{eqnarray} 
\Psi(x_1,x_2;\bar C_{x_1})=\Psi^{\dag}(x_2,x_1;C_{x_1}).
\end{eqnarray}
It is just through these relations that the infinite product of the phase 
factor
on the net of the handled plaquettes (Eq. (2)) should
be equivalent to the original Wilson loop $\Psi(C)$.
To study the property of $\Psi(x,x_0;C)$ under the gauge transformation
\begin{eqnarray}
A'_{\mu}(x)=U(x)A_{\mu}(x)U^{\dag}(x)-\frac{i}{g}\partial_{\mu}U(x)U^{\dag}(x),
\end{eqnarray}
we perform a transformation, $\Psi'(t)=U(x(t))\Psi(t)$,
in Eq. (4). After the rearrangement of the terms we arrive at the 
gauge transformation of $\Psi(x,x_0,C)$:
\begin{eqnarray}
Pexp\left(\int^t_{t_0}dsA'(s)\right)
=U(x(t))Pexp\left(\int^t_{t_0}dsA(s)\right)U^{\dag}(x(t_0)).
\end{eqnarray}

Applying these results to the general case when $A_{\mu}(x)=\sum
\limits^{3}_{a=1}A_{\mu}^a(x)\sigma^a$, with $A_{\mu}^a(x)\neq 0$
for $a=1,2,3$, we first study the behavior of the shifted field strength
$\hat{F}_{\mu\nu}(y,x;C_z)$ in Eqs. (2) and (3). If it can be given as a 
function of the
surface parameters $\{s,t\}$, there is a well-defined ordered surface integral
$Pexp(ig/2\int_S d\sigma^{\mu\nu}\hat{F}_{\mu\nu}(y,z;C_z))$, and the
non-Abelian Stokes theorem (Eq. (1)) will be surely valid. From the definition
of the shifted field strength it is true as long as the
phase factor $\Psi(y,z;C_z)$ can be expressed by the surface parameters 
$\{s,t\}$.

As a matter of fact, however, the phase factor cannot always be
given as the function of $\{s,t\}$
if $A_{\mu}(x)dx^{\mu}$ is not an exact form, i.e. $\partial_{\mu}A_{\nu}(x)
-\partial_{\nu}A_{\mu}(x)=0$ identically.
It is proved as follows. Since the reference point $y$ in Eq. (2) is
arbitrarily chosen on $S$, we can set the origin of the surface coordinate
$\{s,t\}$ at $y$. Then the line integral in Eq. (5) for the phase factor
$\Psi(y,z;C_z)$ is transformed in terms of the surface parameters $\{s,t\}$ to
$$\int^z_yA_{\mu}(\tilde{z})d\tilde{z}^{\mu}=
\int^{(s,t)}_{(0,0)}\left\{A_{\mu}(\tilde{z}(\tilde{s},\tilde{t}))
\frac{\partial\tilde{z}^{\mu}}{\partial\tilde{s}}d\tilde{s}
+A_{\mu}(\tilde{z}(\tilde{s},\tilde{t}))
\frac{\partial\tilde{z}^{\mu}}{\partial\tilde{t}}
d\tilde{t}\right\}.$$
If this line integral is not a path-independent one in the surface coordinate
system $\{s,t\}$, it cannot be given as a function of $\{s,t\}$ and, therefore,
the phase factor $\Psi(z,y;C_z)$ fails to be a function of $\{s,t\}$ too.
Of course, if we choose some special surface coordinates, e.g. that of
a homotopic path family in Ref.[2], the phase factor can be actually expressed
as
a function of the surface parameters, $\Psi(z,y;C_z)=\Psi(s,t)$.
Applying the inverse function theorem in analysis to the transformation,
$x^1=x^1(s,t)$ and $x^2=x^2(s,t)$, we have a locally defined
relations $s=s(x^1,x^2)$ and $t=t(x^1,x^2)$, if
$det(\partial(x^1,x^2)/\partial(s,t))\neq 0$.
Thus $\Psi(s,t)$ can also be given as a function of $\{x^1,x^2\}$:
$\Psi(s,t)=\tilde{\Psi}(x^1,x^2)$. $\tilde{\Psi}(x^1,x^2)$, however,
doesn't satisfy the partial differential equation
\begin{eqnarray}
\partial_{\mu}\tilde{\Psi}(x)=igA_{\mu}(x)\tilde{\Psi}(x),
\end{eqnarray}
unless $A_{\mu}(x)dx^{\mu}$ is an exact form. 
For a non-trivial gauge field this condition doesn't generally hold, and
there is no equivalence between Eqs. (4) and (10).

In the general situation we can say that the flaw with the application of
the non-Abelian Stokes theorem might be in the `handle' phase factor
$\Psi(y,z;C_z)$, which makes the shifted field strength
$\hat{F}_{\mu\nu}(y,z;C_z)$ a path-dependent operator that can't be integrated
with respect to the surface parameters. Actually, even if there is no effect
of the `handle' phase factor, we can still find examples to demonstrate some
ambiguity from the non-Abelian Stokes theorem. To remove the effect of the
`handle' phase factor, we just restrict the generators of the Lie algebra to
its Cartan subalgebras, e.g. $A_{\mu}(x)=A^1_{\mu}(x)\sigma^1$ in SU(2) gauge
theory, then the shifted field strength operator will reduce to field
strength operator $F_{\mu\nu}(x)$.
Our examples are the application of the non-Abelian Stokes theorem to a 
special case
of the gauge invariant operator 
$F_{\mu\nu\rho\sigma}(x,x',y;C_x,C_{x'})=Tr\{\hat{F}_{\mu\nu}(y,x;C_x)
\hat{F}_{\rho\sigma}(y,x';C_{x'})\}$,
the expectation value of which (the correlator of two shifted field
strength) is the basic object of the stochastic vacuum model (SVM)$^{8-10}$.
From Eq. (9) this operator should be gauge invariant under an arbitrary
gauge transformation Eq. (8).
When $x=x'$ and $C=\bar C_{x'}\circ C_{x}$ forms a closed path with the
initial and final point at the same point $x$, the operator reduces to
$$F_{\mu\nu\rho\sigma}(x,x;C)
=Tr\{Pexp\left(-ig\int_Cdz^{\mu}A_{\mu}(z)\right)F_{\mu\nu}(x)
Pexp\left(ig\int_Cdz^{\mu}A_{\mu}(z)\right)F_{\rho\sigma}(x)\}.$$
Obviously, with the help of the non-Abelian Stokes theorem, 
$F_{\mu\nu\rho\sigma}(x,x;C)$ can be rewritten as
\begin{eqnarray}
F_{\mu\nu\rho\sigma}(x,x;C)=Tr\{Pexp\left(-ig/2\int_Sd\sigma^{\mu\nu}\hat{F}_{\mu\nu}(y,z;C_z)\right)
\hat{F}_{\mu\nu}(y,x;C_x)
\\\nonumber
\times
Pexp\left(ig/2\int_Sd\sigma^{\mu\nu}\hat{F}_{\mu\nu}(y,z;C_z)\right)
\hat{F}_{\rho\sigma}(y,x;C_x)\},
\end{eqnarray}
where $y$ is an arbitrary reference point chosen on $S$. 

A specific situation we will study with Eq. (11) is described in
Fig. (1). The contour $\partial S$ here is a rectangular one, 
and $A_{\mu}(x)=A^1_{\mu}(x)\sigma^1$ for each $x\in S$, then the operator 
$F_{\mu\nu\rho\sigma}(x_0,x_0;\partial S)$
is independent of the gauge field on $\partial S$, i.e.
$F_{\mu\nu\rho\sigma}(x_0,x_0;\partial S)=2F^1_{\mu\nu}(x_0)F^1_{\rho\sigma}
(x_0)$.
\begin{figure}
\epsfysize=3in
\hspace{4cm}
\epsffile{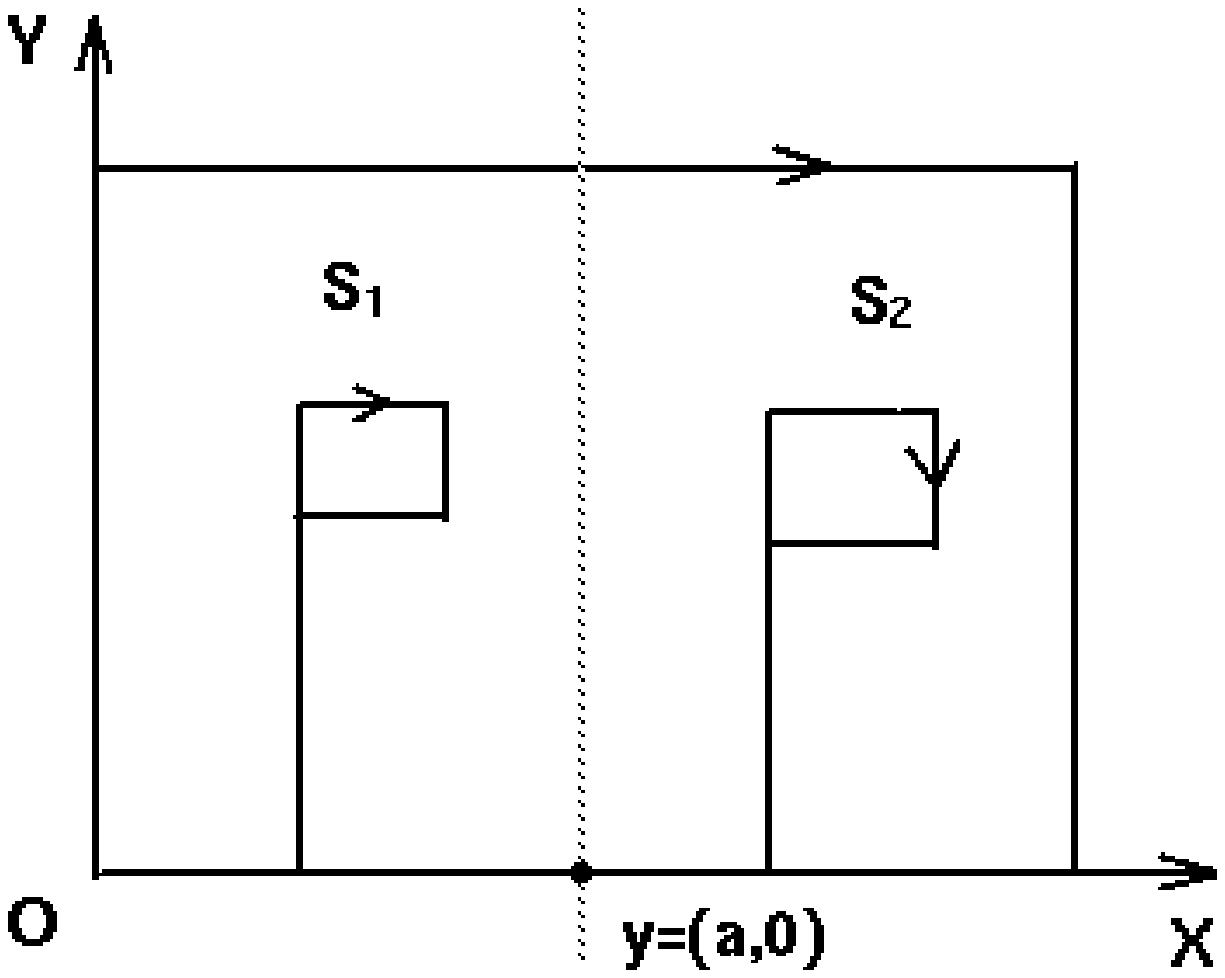}
\caption[]{}
\end{figure}

We apply Eq. (11) to the computation of the operator after 
the following discontinuos gauge transformation on X-Y plane:
\begin{eqnarray}
U(x,y)=exp~(-i\frac{\pi}{4}H(x-a)\sigma^3), 
\end{eqnarray}
where $H(x-a)$ is a Heaviside function, 
$$H(x-a)=\left\{\begin{array}{cc}
0, & x<a, \\
1, & x>a, \\
\end{array}
\right. $$
which transforms the gauge field, $A_{\mu}(x)=A_{\mu}^1(x)\sigma^1$, to 
$$A'_{\mu}(x)=A^1_{\mu}(x)\sigma^1+H(x-a)(A_{\mu}^1(x)\sigma^2-A_{\mu}^1(x)
\sigma^1)
-\frac{1}{g}\frac{\pi}{4}\delta(x-a)\sigma^3{\bf e_x}.$$
Here ${\bf e_x}$ is the unit vector in the direction of X.
The difference will arise if we apply Eq. (11) to the transformed
operator. The reference point here is chosen at $y=(a,0)$,
and the whole $S$ is paved with the net shown in Fig. (1).
From Eqs. (6) and (7), the operator is equivalent to
the product of the handled phase factor on $S$:
\begin{eqnarray}
F_{\mu\nu\rho\sigma}(x_0,x_0;\partial S)=
Tr\{Pexp\left(-ig\int_{L_1}dz^{\mu}A_{\mu}(z)\right)
Pexp\left(-ig\int_{L_2}dz^{\mu}A_{\mu}(z)\right)
\hat{F}_{\mu\nu}(y,x_0)\\\nonumber
\times Pexp\left(ig\int_{L_2}
dz^{\mu}A_{\mu}(z)\right)Pexp\left(ig\int_{L_1}dz^{\mu}A_{\mu}(z)\right)
\hat{F}_{\rho\sigma}
(y,x_0)\},
\end{eqnarray}
where $Pexp\left(ig\int_{L_i}dz^{\mu}A_{\mu}(z)\right)$, $i=1,2$, is the product of 
the phase factor on the net of the handled plaquettes on $S_i$ ($i=1,2$),
and is equal to
$Pexp\left(ig/2\int_{S_i}d\sigma^{\mu\nu}\hat{F}_{\mu\nu}(y,z;C_{z})\right)$,
respectively. 
After the gauge transformation Eq. (12), there
are 
$$\hat{F'}_{\mu\nu}(y,x_0)=F^1_{\mu\nu}(x_0)\sigma^1,$$
$$\hat{F'}_{\mu\nu}(y,z)=F^1_{\mu\nu}(z)\sigma^1$$ on $S_1$, and
$$\hat{F'}_{\mu\nu}(y,z)=F^1_{\mu\nu}(z)\sigma^2$$
on $S_2$.
Substituting these results into Eq. (11), we obtain
\begin{eqnarray}
F'_{\mu\nu\rho\sigma}(x_0,x_0;C)=Tr\{
Pexp\left(-ig/2\int_{S_2}d\sigma^{\mu\nu}
F^1_{\mu\nu}(z)\sigma^2\right)F^1_{\mu\nu}(x_0)\sigma^1 \\\nonumber
\times Pexp\left(ig/2\int_{S_2}d\sigma^{\mu\nu}F^1_{\mu\nu}(z)\sigma^2\right)
F^1_{\rho\sigma}(x_0)\sigma^1\}.
\end{eqnarray}
It is obviously not equal to the previous result because
it involves the gauge field at other points than $x_0$ and,
therefore, an ambiguity of the gauge invariance of the operator 
$F_{\mu\nu\rho\sigma}(x_0,x_0;\partial S)$ will arise unless
the field is a pure gauge one, i.e. the field strength $F_{\mu\nu}(z)$ 
vanishes identically on $S$.

Furthermore, we can find an example that directly shows the
ambiguity of the non-Abelian Stokes theorem in the computation of the Wilson
loop, when the generators of the Lie algebra are still confined in its
Cartan subalgebras.
Here we adopt polar coordinate (Fig. 2). 
In this case
the gauge field on $S$ is given as a discontinuous one:
$$A_{\mu}(x)=\left\{ \begin{array}{cc}
A_{\mu}^1(r,\theta)\sigma^1, & r>r_2~ or~ r<r_1, \\
\frac{1}{g}\frac{\pi}{4}\partial_{r}F(r){\bf e_r}\sigma^3, & r_1<r<r_2. 
\end{array}
\right. $$
The smooth function $F(r)$ is defined as
$$ F(r)=\int^{r_2}_{r}f(r)dr/\int^{r_2}_{r_1}f(r)dr,$$
where the function $f(r)$ is given as
$$f(r)=\left\{ \begin{array}{cc}
exp(-\frac{1}{r-r_1}-\frac{1}{r-r_2}), &    r_1<r<r_2, \\
0, &    elsewhere.
\end{array}   
\right.  $$
\newpage
Such a smooth function has the following property:
$$F(r)=\left\{\begin{array}{cc}
1, & r\leq r_1, \\
0, & r\geq r_2, \\
decrease~ from~1~to~0, & r_1<r<r_2.
\end{array}
\right. $$

\begin{figure}
\epsfysize=3in
\hspace{4cm}
\epsffile{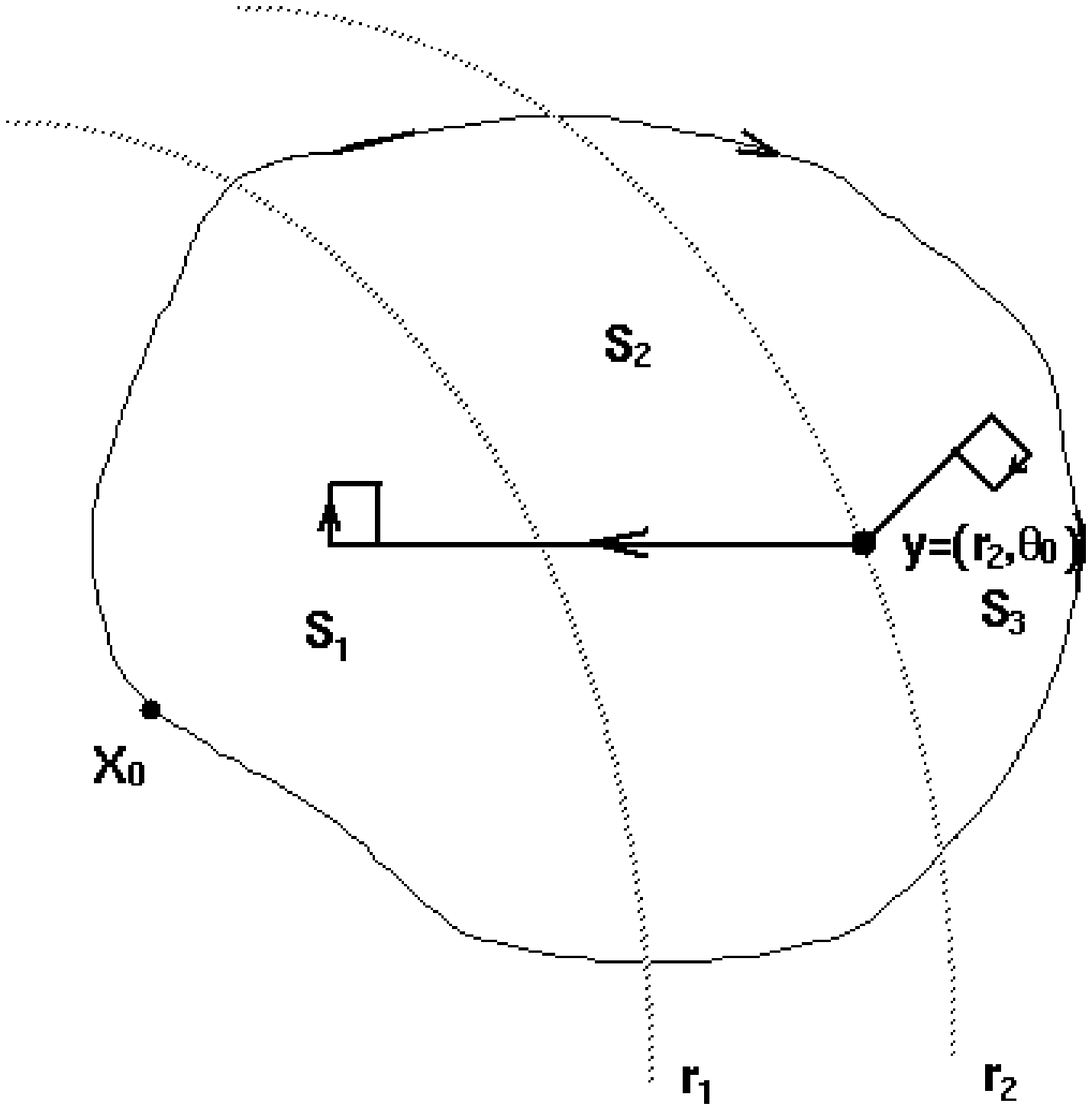}
\caption[]{}
\end{figure}

\noindent
The whole surface $S$ is the union of the three parts: $S=
S_1\cup S_2\cup S_3$.
According to the original definition of the operator
$F_{\mu\nu\rho\sigma}
(x_0,x_0;\partial S)$, it is given as follows:
\begin{eqnarray}
F_{\mu\nu\rho\sigma}(x_0,x_0;\partial S)=
Tr\{F^1_{\mu\nu}(x_0)\sigma^2Pexp\left(ig\int_{C_3}A_{\mu}(z)\sigma^1dz^{\mu}\right)
\\\nonumber
\times F^1_{\rho\sigma}(x_0)\sigma^2Pexp\left(-ig\int_{
C_3}A_{\mu}(z)\sigma^2dz^{\mu}\right)\},
\end{eqnarray}
where the path $C_3$ is $\partial S_3$ excluding the arc $r_2\cap S$.
The operator is determined by
the gauge field $A_{\mu}(x)$ on $C_3$,
since the phase factor between $r=r_1$ and
$r=r_2$, $Pexp~(\int^{r_2}_{r_1}
i\frac{\pi}{4}\partial_{r}F(r)\sigma^3dr)=exp~(-i\frac{\pi}{4}\sigma^3)$,
rotates the field strength operator $F_{\mu\nu}^1(x_0)\sigma^1$ at $x_0$ to 
the direction of $\sigma^2$ in the internal space.

To apply the non-Abelian Stokes theorem to the operator, we choose the 
reference point at $(r_2,\theta_0)$. Then we have
$$\hat{F}_{\mu\nu}(y,x_0)=exp~(-i\frac{\pi}{4}\sigma^3)
F_{\mu\nu}^1(x_0)\sigma^1exp~(i\frac{\pi}{4}\sigma^3)
=F^1_{\mu\nu}(x_0)\sigma^2,$$
and
$$Pexp~(ig/2\int_S \hat{F}_{\mu\nu}(y,z)d\sigma^{\mu\nu})
=Pexp~(ig/2\int_{S_{12}}F^1_{\mu\nu}(z)\sigma^2
d\sigma^{\mu\nu})$$
$$\times Pexp~(ig/2
\int_{S_3}F^1_{\mu\nu}(z)\sigma^1d\sigma^{\mu\nu})
Pexp~(ig/2\int_{S_{11}}F^1_{\mu\nu}(z)\sigma^2d\sigma^{\mu\nu}),$$
where $S_1=S_{11}\cup S_{12}$ with respect to the ordering of the plaquettes
on the net (Fig. 2),
since the contribution from $S_2$, a area with pure gauge, is zero.
After substituting the contributions from the three different parts into 
Eq. (11),
and considering the fact:
$$Pexp~(ig/2\int_{S_3}F^1_{\mu\nu}(z)\sigma^1d\sigma^{\mu\nu})
=Pexp~(ig\int_{\partial S_3}A^1_{\mu}(z)\sigma^1dz^{\mu}),$$
we have
\begin{eqnarray}
F_{\mu\nu\rho\sigma}(x_0,x_0;C)&=&Tr\{ Pexp\left(-ig\int_{\partial S_3}
A_{\mu}^1(z)\sigma^1 dz^{\mu}\right)F^1_{\mu\nu}(x_0)\sigma^2\\\nonumber
&\times& Pexp\left(ig\int_{\partial S_3}A^1_{\mu}(z)\sigma^1dz^{\mu}\right)
F_{\rho\sigma}^1(x_0)\sigma^2\}.
\end{eqnarray}
The difference between the results given in Eq. (15) and Eq. (16) is
an additional phase factor $Pexp~(ig\int_{r'_2}A_{\mu}(z)dz^{\mu})$,
where $r'_2=r_2+\epsilon$ ($\epsilon$ is an infinitesimal positive number), 
in the phase 
factor $Pexp~(ig\int_{\partial_{S_3}}
A_{\mu}(z)dz^{\mu})$ in Eq.(16), because $\partial S_3=r'_2\circ C_3$.

It is clearer to see the difference if we choose a new gauge by performing
a gauge transformation,
$U(r,\theta)=exp(-i\frac{\pi}{4}F(r)\sigma^3)$, over $S$.
The gauge field on $S$ will therefore be transformed to
$$A_{\mu}(x)=\left\{\begin{array}{cc}
A_{\mu}(r,\theta)\sigma^2, & r<r_1, \\
A_{\mu}(r,\theta)\sigma^1, & r>r_2, \\
0, & r_1<r<r_2.
\end{array}
\right. $$
Then it is easier to compute
the shifted field
strength operators, $\hat{F}_{\mu\nu}(y,z)$, on
the three different parts of $S$ and reproduce the results in Eq. (15) and 
Eq. (16).
The same operator $F_{\mu\nu\rho\sigma}(x_0,x_0;\partial S)$, with the
concerned Wilson
loop treated differently
as a `one-dimensional object' or a `two-dimensional object' through
the non-Abelian Stokes theorem, will be given two different
results. This example demonstrates the ambiguity from the transformation of the line integral to
the surface integral in Eq. (1), even if the effect of the `handle' phase
factor is removed.

Finally we should mention a problem in lattice gauge field theory that is
related to our discussion. In the literature available the non-Abelian
Stokes theorem is widely used to the expansion of a plaqutte operator 
with respect to lattice spacing $a$,  
which is crucial in the construction of the lattice actions with the lattice 
artifact removed
to a certain order of the lattice spacing $a$ (improved action approach). A 
plaquette operator $U_{\mu\nu}$ is expanded according to the theorem as 
follows$^{11}$:
\begin{eqnarray}
U_{\mu\nu}=U_{\mu}U_{\nu}U^{\dag}_{\mu}U_{\nu}^{\dag}
=Pexp\left(a^2\int^1_0ds\int^1_0dt\hat{F}_{\mu\nu}
(x+as\hat{\mu}+at\hat{\nu})\right)\\\nonumber
=I+\sum \limits_{n=1}^{\infty}\prod
\limits^n_{i=1}\int^{s_{i-1}}_0ds_i\int^1_0dt_ia^2\hat{F}_{\mu\nu}
(x+as_i\hat{\mu}+at_i\hat{\nu})\times\cdots a^2(x+as_n\hat{\mu}+at_n\hat{\nu}).
\end{eqnarray}
Then one needs to perform the Taylor expansion of
$\hat{F}(x+as\hat{\mu}+at\hat{\nu})$ around $x$ with respect to $(s,t)$
with the help of the relation
\begin{eqnarray}
\partial^n_{\mu}\partial^m_{\nu}\hat{F}_{\mu\nu}(x)
=D^n_{\mu}D^m_{\nu}F_{\mu\nu}(x).
\end{eqnarray}
In doing so, it is taken for granted that the shifted field strength operator,
$$\hat{F}_{\mu\nu}(x+as\hat{\mu}+at\hat{\nu})=\Psi(s,t)
F_{\mu\nu}(x+as\hat{\mu}+at\hat{\nu})\Psi^{\dag}(s,t),$$
is a function of $(s,t)$ as the field strength operator
$F_{\mu\nu}(x+as\hat{\mu}+at\hat{\nu})$ itself. Our previous discussions prove
that it is not generally valid for the situation of a non-trivial
gauge field and, moreover, Eq. (18) is true only when $A_{\mu}(x)dx^{\mu}$ is 
an exact form, i.e. there is the relation Eq. (10).  

Of course a plaquette operator can also be expanded by means of
Baker-Hausdorff formula or the choice of axial
gauge$^{12-13}$, e.g.
by imposing
$$A_1(x_1,x_2)=0,$$
$$A_2(0,x_2)=0,$$
on the field configuration, which simplifies the plaquette operator
considerably.
However, for a gauge field in the general situation,
i.e. $A_{\mu}(x)\neq 0$ ($\mu=1,2$) in any of a gauge we choose, the
non-Abelian Stokes theorem is the only effective tool for a
convenient expansion of a plaquette operator. With the problems in its
application we
have discussed,
it should be taken an approximate
rather than an exact approach if we are dealing with a non-trivial gauge field.

\vspace{6mm}

{\bf ACKNOWLEDGMENTS}. 
The work is partly supported by National Nature
Science Foundation of China under Grant 19677205.

\end{document}